# Stability of xenon-sodium compounds at moderately low pressures

*Shaoxiong Wang‡[1], Junhao Peng‡[1], Huafeng Dong[1,2]\*, Minru Wen[1], Xin Zhang[1], Fugen Wu[2,3]*

[1] School of Physics and Optoelectronic Engineering, Guangdong University of Technology, Guangzhou 510006, China

[2] Guangdong Provincial Key Laboratory of Information Photonics Technology, Guangdong University of Technology, Guangzhou 510006, China

[3] School of Materials and Energy, Guangdong University of Technology, Guangzhou 510006, China

\* Corresponding Author: Huafeng Dong hfdong@gdut.edu.cn,

‡These authors contributed equally.

KEYWORDS: Xenon, Sodium, Structure prediction

**Abstract**

A growing body of theoretical and experimental evidence suggests that inert gases (He, Ne, Ar, Kr, Xe, Rn) become less and less inert under increasing pressure. Here we use the ab initio evolutionary algorithm to predict stable compounds of Xe and Na at pressures below 100 GPa, and find three stable compounds, NaXe, NaXe$_3$ and NaXe$_4$. The NaXe belongs to a well-known cubic CsCl structure type. The NaXe$_4$'s structure is common in amphiboles, whereas the NaXe$_3$ has a unique structure, analogous to the "post-perovskite" orthorhombic CaIrO$_3$-type structure with Ir atoms removed. This is the first time that a cation-vacant version of the CaIrO$_3$ is found in any compound. NaXe, NaXe$_3$ and NaXe$_4$ are found to be metallic.

**Introduction**

Inert (or "noble") gases have closed valence shell - it is difficult to remove from or add to it an electron, or to share electrons with other atoms. This inertness decreases from light (He, Ne) to heavy (Xe, Rn) noble gases, and with increasing pressure. One should distinguish four situations:

(1) Van der Waals compounds, which are stabilized under pressure because of denser packing of two (or more) atoms of different size. For example, at the pressure of 4.78 GPa, Somayazulu discovered a stable Xe-H compound Xe(H$_2$)$_8$ [1].

(2) Neutral insertion of noble gases into ionic or electride structures. Examples are Na$_2$He and Na$_2$OHe [2] and such compounds as CaF$_2$He and MgF$_2$He [3].

(3) Bonding with electronegative elements. For example, Xe forms stable fluorides already at ambient pressure [4], and stable oxides at pressures >74 GPa [5, 6] – whereas He would require terapascal pressures to form stable compounds with F or O. There is evidence for incorporation of Xe into quartz (SiO$_2$) at rather low pressures (0.7-5 GPa) and elevated temperatures (500-1500 K) [7].

(4) Bonding with electropositive elements. For example, Xe forms stable compounds with Fe and Ni at 150–350 GPa [8].

Here we consider the latter situation, where a noble gas forms a stable compound with a highly electropositive element. Large electronegativity difference should stabilize such compounds. On the other hand, Xe should form such compounds more easily, as its outer electrons are more polarizable, i.e., more susceptible to external influence and more prone to chemical bonding. Indeed, at relatively low pressures of >43 GPa, we predict the stability of NaXe, NaXe$_3$ is predicted to appear at pressures >63 GPa and NaXe$_4$ is predicted to appear at pressures > 82 GPa.

**Results and discussion**

Phase diagram and new structures. Figure 1 (a) shows the phase diagram of the Na-Xe system in the pressure range 0-100 GPa. $Pm\bar{3}m$-NaXe become stable at 43 GPa, *Cmcm*-NaXe$_3$ become stable at 63 GPa and *C*2/*m*-NaXe$_4$ become stable at 82 GPa, respectively, and remain stable at least up to 100 GPa (Figure S1-S3). Our phase diagram also shows predicted transitions in elemental Na (bcc to fcc), which has been reported before[9, 10], and the phase transition in Xe (fcc to hcp), which is also known from experiments [11, 12]. Figure 1b shows that the enthalpy of formation of NaXe, NaXe$_3$ and NaXe$_4$ become much more negative with pressure, i.e., these compounds become much more exothermic. One can also see that other stoichiometries, namely Na$_2$Xe$_{15}$, Na$_2$Xe$_9$, NaXe$_2$, Na$_2$Xe, Na$_9$Xe, Na$_7$Xe$_3$ and Na$_2$Xe$_7$, come close to the convex hull (Figure 1b) and might become stable at pressures higher than 100 GPa. Phonon calculations for $Pm\bar{3}m$-NaXe,

*Cmcm*-NaXe₃ and *C*2/*m*-NaXe₄ show no imaginary frequencies, indicating that compounds are dynamically stable in the pressure range of their stability. At ambient pressure all structures have imaginary phonon frequencies (Figures S4-S9), and are therefore not quenchable to ambient conditions.

Structural features and the origin of metallicity. Structural parameters of $Pm\bar{3}m$-NaXe, *Cmcm*-NaXe₃ and *C*2/*m*-NaXe₄ are listed in Table 1. $Pm\bar{3}m$ -NaXe has a well-known CsCl-type cubic structure (Figure 2a), where atoms of both types have 8-fold coordination. *Cmcm*-NaXe₃ has a very unusual structure (Figure 2c), which can be derived from the CaIrO₃-type ("post-perovskite") structure (Figure 2d) by removing the small cation (Ir). The CaIrO₃-type phase of MgSiO₃ ("post-perovskite") is universally believed to be the main mineral phase of the Earth's lowermost mantle (the D" layer)[13]. All known post-perovskite phases have ABX₃ stoichiometry; *Cmcm*-NaXe₃ is the first post-perovskite phase with stoichiometry BX₃.

Cation-deleted versions of the perovskite structure are well known – deleting the large cation, one obtains the ReO₃ structure type. Deleting the small cation, one obtains the Cu₃Au structure, which is nothing but an ordered version of the cubic close packed structure. Cation-deleted versions of the post-perovskite structure were not known to us before, but *Cmcm*-NaXe₃ gives the first example of such a structure, with the deletion of the small cation. In this structure, each Na atom is coordinated by 8 Xe atoms.

Calculated electronic band structures (Figure 3) show that $Pm\bar{3}m$ -NaXe, *Cmcm*-NaXe₃ and *C*2/*m*-NaXe₄ are metallic: in NaXe three bands, in NaXe₃ four bands and in NaXe₄ four bands cross the Fermi level. While NaXe has an odd number of electrons per unit cell and therefore has have half-filled bands, NaXe$_3$ and NaXe$_4$ have an even number of electrons and could be insulators, yet are also metals. They all have band structure similar to n-doped semiconductors: there is a gap (~2.7 eV) between fully occupied and partially filled bands. The latter can be viewed as electrons brought by the Na atoms and donated to whole crystal.

In Figure 4 we compare the calculated electronic densities of states of $Pm\bar{3}m$-NaXe (43 GPa), *Cmcm*-NaXe$_3$ (63 GPa) and *C*2/*m*-NaXe$_4$ (82 GPa) with the densities of states obtained by removing Na atoms. One can see that removal of Na atoms makes the structures insulating, but we also see a considerable interaction of Na and Xe electrons in the energy range of [E$_F$-2 eV, E$_F$] eV (E$_F$: Fermi level).

In order to further explore the distribution of electrons that make the structure metallic, we calculated the partial charge densities of three structures (Figure 5) in the energy range of [E$_F$-2 eV, E$_F$]. The charge is mainly distributed between atoms. In $Pm\bar{3}m$-NaXe (43 GPa), the charge is mainly distributed between Xe atoms; In *Cmcm*-NaXe$_3$ (63 GPa) and *C*2/*m*-NaXe$_4$ (82 GPa), the charge is mainly distributed on the connecting line of adjacent Na atoms. Although these electrons are jointly provided by Na and Xe atoms, we are curious about the main source of these electrons, thus we performed Bader analysis (Table 2) [14-17]. This analysis shows that each Na atom loses 0.69 electrons in $Pm\bar{3}m$-NaXe, 0.74 electrons in *Cmcm*-NaXe$_3$ and 0.76 electrons in *C*2/*m*-NaXe$_4$.

**Conclusions**

By employing the evolutionary algorithm USPEX, we have predicted new stable compounds in the Na-Xe system in the pressure range 0-100 GPa: $Pm\bar{3}m$-NaXe (stable at >43 GPa and up to at least 100 GPa), $Cmcm$-NaXe$_3$ (stable at >63 GPa and up to at least 100 GPa) and $C2/m$-NaXe$_4$ (stable at >82 GPa and up to at least 100 GPa). NaXe$_3$ has a unique structure related to post-perovskite CaIrO$_3$-type structure. Electronic structure calculations suggest that $Pm\bar{3}m$-NaXe, $Cmcm$-NaXe$_3$ and $C2/m$-NaXe$_4$ are metals. According to Bader analysis, electrons in the sodium atoms are transferred to the Xe atoms. The Xe atoms do not behave as fully inert; instead, the electrons of Na and Xe are coupled, and we see Xe atoms polarizing towards neighboring Na atoms in $Cmcm$-NaXe$_3$. Despite becoming stable at relatively low pressures, NaXe, NaXe$_3$ and NaXe$_4$ will not survive decompression to ambient pressure.

**Methods**

The evolutionary algorithm USPEX[18] is used here to predict new stable structures. It searches for the lowest enthalpy structure under pressures of 0 GPa, 50 GPa, and 100 GPa, and can predict stable compounds and the element ratio of the structure. A number of applications illustrate its power [5, 9, 18, 19]. In order to make the search more thorough, after obtaining the element ratio of the structure, continue to search for the fixed composition. Structure relaxations were performed using Perdew–Burke–Ernzerhof (PBE) functional[20] in the framework of the projector augmented wave (PAW) method[21], as implemented in the VASP code[22]. For Na atoms we used PAW potentials with 1.4 a.u. core radius and $2s^2 2p^6 3s^1$ electrons treated as valence. For Xe the core radius was 1.45 a.u. and $4s^2 4p^6 4d^{10} 5s^2 5p^6$ electrons were treated as valence. We use 500 eV plane-wave kinetic energy cutoff, total energy convergence criterion is set to $10^{-8}$ eV, and the structure parameters have fully optimized. The first-generation structures are randomly created, all structures relax under constant pressure and 0 K, and enthalpy is used as the fitness. The highest energy structures (40%) were discarded and a new generation was created, 30% of which were random and 70% from the lowest enthalpy structure through heredity, lattice mutation and transmutation. In addition, in order to determine the thermodynamically stable structure, the normalized enthalpy of formation of the structure was calculated:

$$Ef = \frac{(Total(Na_x Xe_y) - xTotal(Na) - yTotal(Xe))}{(x+y)}$$

. Then, using these enthalpies of formation, the convex hull was constructed – a phase is deemed thermodynamically stable if it lies on the convex hull, i.e., is lower in enthalpy than any isochemical assemblage of other phases. In order to judge the dynamical stability of the structure, the phonon spectrums of $Pm\bar{3}m$-NaXe (0 GPa and 43 GPa), $Cmcm$-NaXe$_3$(0GPa and 63 GPa) and $C2/m$-NaXe$_4$ (0 GPa and 82GPa) were calculated with PHONOPY code [23]. Bader charge analysis and local charge density were used to analyze the charge transfer near the Fermi level [14-17]. VASPKIT code is used in data processing [24].

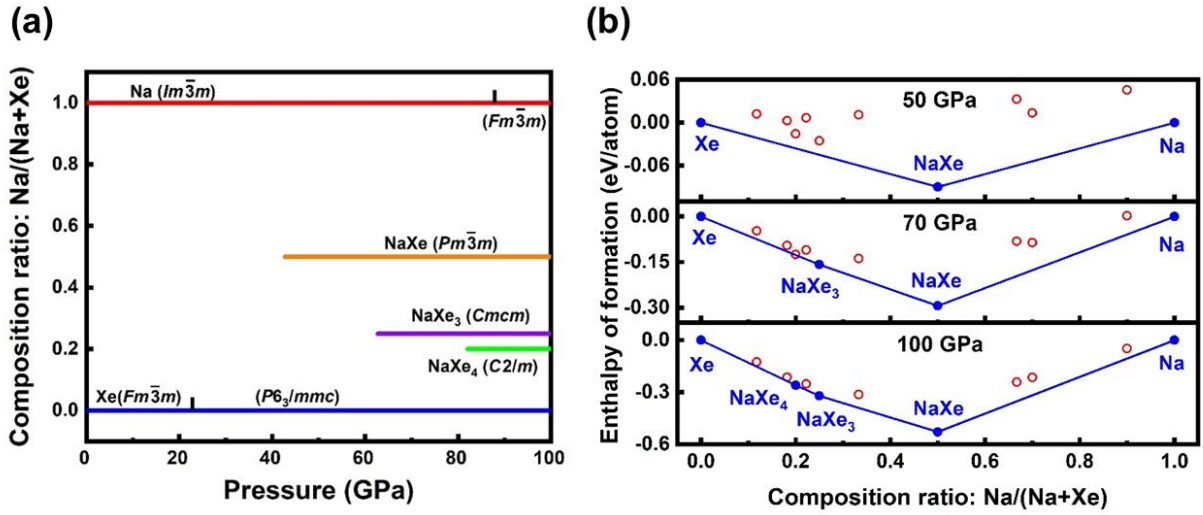

Figure 1. Phase stability in the Na-Xe system. (a) Phase diagram of the Na-Xe system at the pressure of 0-100 GPa at temperature of 0 K. (b) Convex hulls at pressures of 50, 70, 100 GPa.

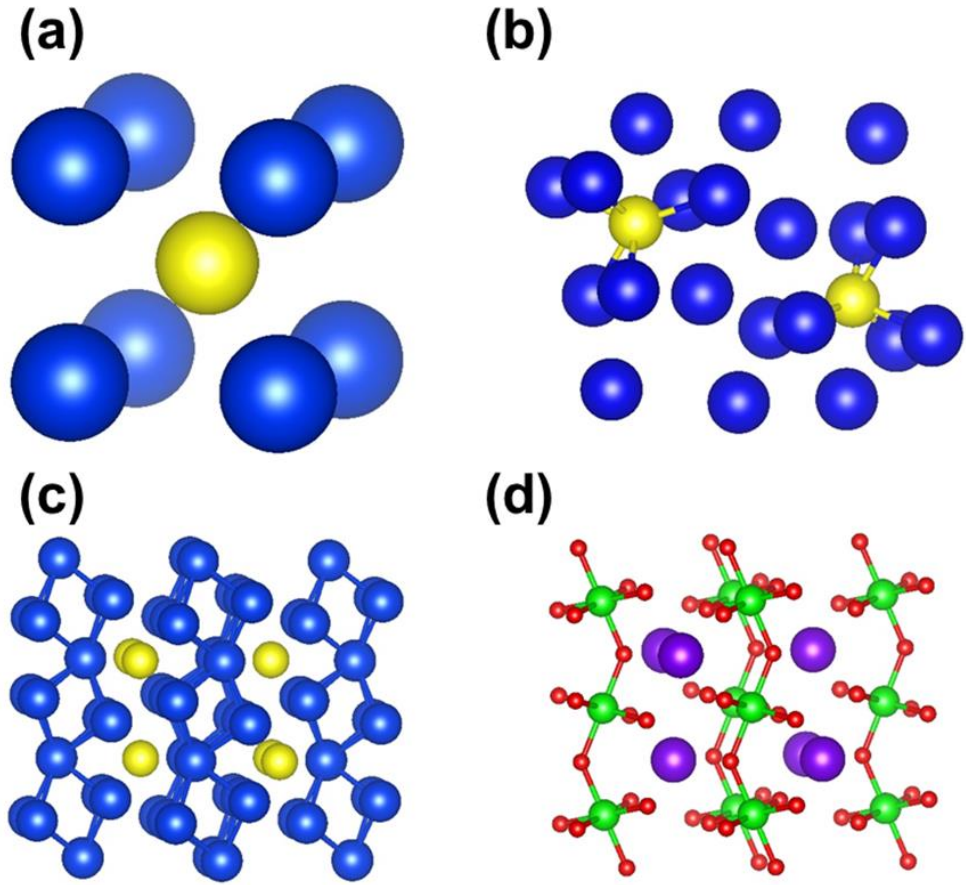

Figure 2. Crystal Structures of Na-Xe compounds. (a) $P m\bar{3}m$ -NaXe, the center of the cube is Na, and the eight atoms around it are Xe;(b) $C2/m$-NaXe$_4$, the blue ball is Xe and the yellow ball is Na; (c) $Cmcm$-NaXe$_3$, the blue ball is Xe and the yellow ball is Na; (d) CaIrO$_3$ (post-perovskite phase), in which purple, red and cyan spheres are Ca, O and Ir atoms, respectively.

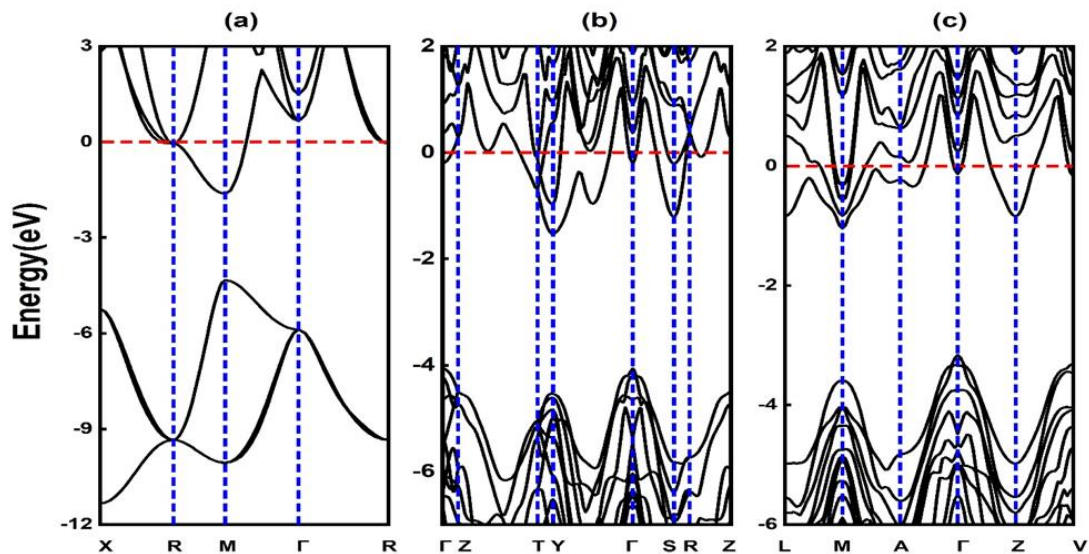

Figure 3. Band structure of the new structures. (a) Band structure of $Pm\bar{3}m$－NaXe at the pressure of 43 GPa；(b) Band structure of *Cmcm*-NaXe$_3$ at the pressure of 63 GPa; (c) Band structure of *C*2/*m*-NaXe$_4$ at the pressure of 82 GPa.

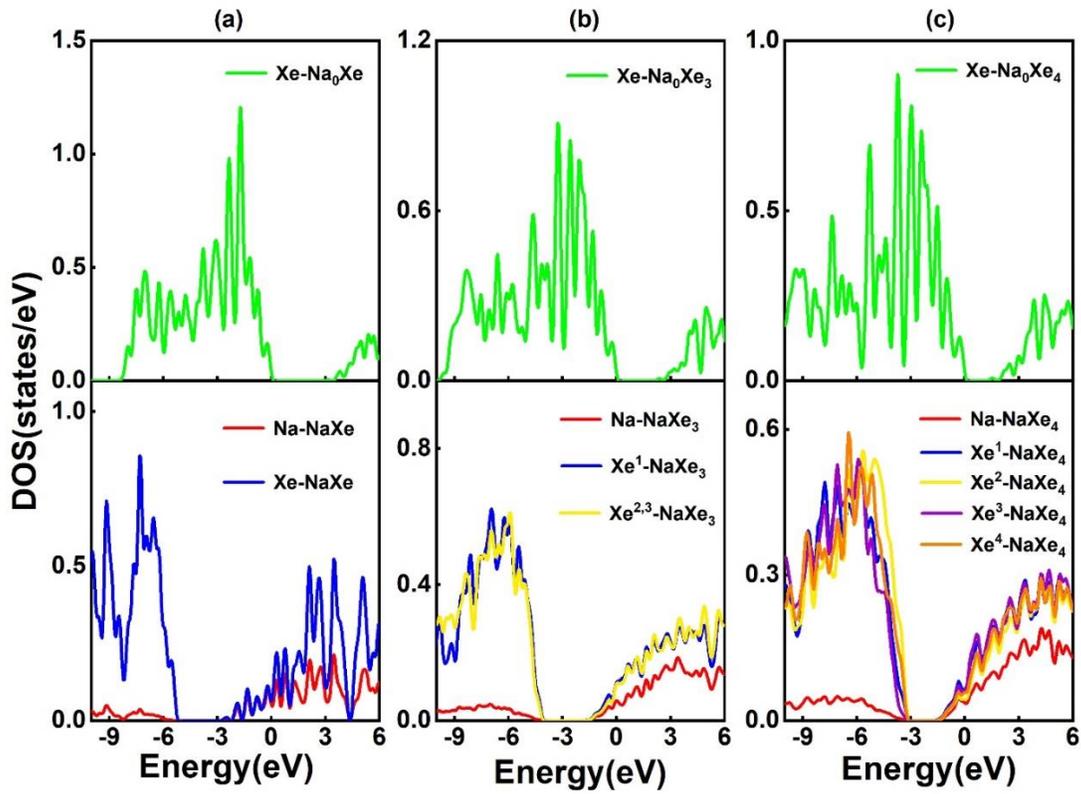

Figure 4. Density of states of the new structures. (a) Density of states of $Pm\bar{3}m$-NaXe at 43 GPa; (b) Density of states of $Cmcm$-NaXe$_3$ at 63 GPa; (c) Density of states of $Cmcm$-NaXe$_4$ at 82 GPa. The density of states of elemental Xe under corresponding pressure is also added.

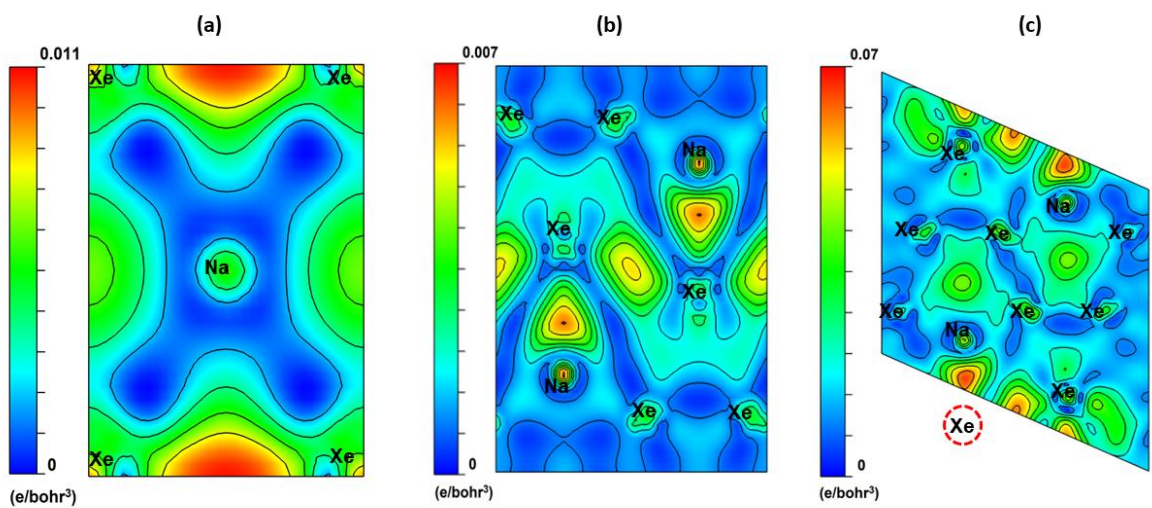

Figure 5. Partial charge density of three structures. The electron density is expressed in e/Å$^3$. The sections of $Pm\bar{3}m$-NaXe (43 GPa) and $Cmcm$-NaXe$_3$ (63 GPa) are (1 1 0), and the section of $C2/m$-NaXe$_4$ (82 GPa) is (- 1 1 0).

Table 1. Lattice parameters and atomic coordinates for $Pm\bar{3}m$ -NaXe (at 43 GPa), $Cmcm$-NaXe$_3$ (at 63 GPa) and $C2/m$-NaXe$_4$ (at 83 GPa).

| Structure | Phase | Parameters (Å, degree) | Wyckoff Position | | | | |
|---|---|---|---|---|---|---|---|
| | | | atom | Site | x | y | z |
| NaXe | $Pm\bar{3}m$ | a=b=c=3.225 | Na | b | 0.500 | 0.500 | 0.500 |
| | | α=β=γ=90 | Xe | a | 0.000 | 0.000 | 0.000 |
| NaXe$_3$ | $Cmcm$ | a=b=5.83 c=8.27 α=β=90 γ= 147.92 | Na | c | 0.761 | 0.240 | 0.750 |
| | | | Xe | c | 0.449 | 0.551 | 0.750 |
| | | | Xe | f | 0.135 | 0.865 | 0.556 |
| NaXe$_4$ | $C2/m$ | a=b= 5.63 c= 11.43 α=β= 112.25 γ= 32.65 | Na | i | 0.821 | 0.821 | 0.692 |
| | | | Xe | i | 0.866 | 0.866 | 0.305 |
| | | | Xe | i | 0.180 | 0.180 | 0.070 |
| | | | Xe | i | 0.384 | 0.384 | 0.543 |
| | | | Xe | i | 0.493 | 0.493 | 0.158 |

Table 2 | Bader analysis

### $Pm\bar{3}m$-NaXe (43 GPa) Charge transfer situation

| Atom | Charge |
| --- | --- |
| Na | +0.69 |
| Xe | -0.69 |

### $Cmcm$-NaXe$_3$ (63 GPa) Charge transfer situation

| Atom | Charge |
| --- | --- |
| Na | +0.74 |
| Xe | -0.24 |
| Xe | -0.25 |
| Xe | -0.25 |

### $C2/m$-NaXe$_4$ (82 GPa) Charge transfer situation

| Atom | Charge |
| --- | --- |
| Na | +0.76 |
| Xe | -0.22 |
| Xe | -0.12 |
| Xe | -0.24 |
| Xe | -0.18 |

**Author Contributions**

Huafeng Dong conceived the project. Shaoxiong Wang conducted the *ab initio* calculations, analyzed the obtained results and wrote the paper. Junhao Peng guided the computational method and proposed the paper modification. Huafeng Dong, Shaoxiong Wang, Junhao Peng, Minru Wen, Xin Zhang, Fugen Wu contribute to the discussion of the manuscript. All authors have given approval to the final version of the manuscript.

**Acknowledgements**

This work is supported by the National Natural Science Foundation of China (Grant No. 11604056, and 11804057) and the special fund for climbing program of Guangdong Province, China. We thank the Center of Campus Network & Modern Educational Technology, Guangdong University of Technology, Guangdong, China for providing computational resources and technical support for this work. Many thanks to Prof. Artem R. Oganov, Prof. Alexander F. Goncharov and Dr. Lin Wang for their useful discussions.